\documentstyle[psfig]{l-aa}

\begin{document}

\thesaurus {06 (08.02.2, 08.02.4, 08.09.2 (SS Lac), 10.15.2 (NGC 7209))}

\title{Spectroscopic orbit of the ex-eclipsing binary SS Lac \\
in the young open cluster NGC 7209}

\author{
       Lina Tomasella\inst{1}
\and   Ulisse Munari\inst{2}
       }
\offprints{U.Munari}

\institute {
Osservatorio Astrofisico del Dipartimento di Astronomia, 
Universit\`a di Padova, I-36012 Asiago (VI), Italy
\and
Osservatorio Astronomico di Padova, Sede di Asiago, 
I-36012 Asiago (VI), Italy
}
\date{Received date..............; accepted date................}

\maketitle

\markboth{L.Tomasella and U.Munari: Spectroscopic orbit of the ex-eclipsing
binary SS~Lac}{L.Tomasella and U.Munari: Spectroscopic orbit of the ex-eclipsing
binary SS~Lac} 

\begin{abstract}
The no-longer-eclipsing system SS~Lac in the young open cluster NGC 7209 has
been recently announced to show a constant radial velocity.  Puzzled by this
finding, we have monitored the system during 1997 obtaining 24 Echelle+CCD
spectra over 8 orbital revolutions. Our spectra reveal a nice orbital motion
with periodic splitting and merging of spectral lines from both components.

An accurate orbit has been derived, together with individual masses of the
stars. SS~Lac presents a moderately eccentric orbit and a probably full 
synchronization between stellar rotation and orbital revolution.

\end{abstract}
\maketitle

\section{Introduction}

According to recent photometry (cf. Vansevicius et al. 1997) SS~Lac belongs
to the young open cluster NGC 7209 . Hoffmeister (1921), Dugan \& Wright
(1935), Wachmann (1936), Nekrasova (1938) and Kordylewski et al. (1961)
described SS~Lac as an eclipsing binary of 14.4 days period, with
significant eccentricity because the equal depth secondary eclipse occurred
at phase 0.57.  However, more recent observations by Zakirov \& Azimov
(1990), Lacy (1992), Mossakovskaya (1993) and Schiller et al. (1996) show
that the eclipses no longer occur. The end of the eclipsing phase is set
around 1940 by Mossakovskaya (1993) and around 1960 by Lehmann (1991). The
latter ascribes the rotation of the plane of the orbit to the presence of an
unseen third body in the system, orbiting at a great distance the closer
central pair.

Quite puzzling has been the report by Schiller et al. (1991) and Schiller \&
Milone (1996) that 1983-84 spectra of SS~Lac did not reveal the expected
double--lined nature of the spectra. Very little, if any, variation in
radial velocity was observed. Using available predictions for the rate of
change of the SS~Lac orbital inclination, Schiller \& Milone (1996) expected
a semi-amplitude of the radial velocity curve of 150 km sec$^{-1}$, far in
excess of their instrumental resolution. They suggested that SS~Lac could be
a triple system suddenly become chaotic or that a close encounter with
another NGC 7209 member could have ionized the binary (cf. Schiller 1996).

Stimulated by the Schiller et al. (1991) report, Etzel et al. (1996),
Stefanik et al. (1996) and Etzel \& Volgenau (1996) announced in IAU
Circulars that, according to their preliminary spectroscopic observations,
SS~Lac is a double--lined system, with indications of variability in the
radial velocities. So far no investigation of the orbital motion of SS~Lac
has appeared in literature.

As an ex-eclipsing binary in a well populated young open cluster, SS~Lac
clearly deserved further investigations to clarify the whole issue. In this
note we report about our spectroscopic monitoring of SS~Lac performed to the
aim of confirming the binary nature and to derive the spectroscopic orbit.

\section{Observations}

Twenty-four Echelle+CCD spectra of SS~Lac have been obtained with the Asiago
1.82 m telescope during 1997. They cover the range from 4300 to 6600 \AA\
with a resolution of $\lambda/\bigtriangleup\lambda \sim$22,500 at H$\beta$
(from the FWHM of comparison spectrum Thorium lines). The spectra have been
reduced in a standard way with the IRAF software package. A sample of
H$\alpha$ profiles recorded on our spectra is shown in Figure~1, where the
periodic splitting and merging of the individual components is evident.

On our spectra we measured only the radial velocities of the H$\alpha$,
H$\beta$ and H$\gamma$ lines (the latter over two distinct Echelle orders).
A few other metallic lines were indeed visible, but too weak for meaningful
measurements. The radial velocities reported in Table~1 come from the
hydrogen lines, weighted according to the S/N of the stellar continuum
around the line (typically 6:4:3 for H$\alpha$:H$\beta$:H$\gamma$, with
an average S/N=50--60 around H$\alpha$).

\begin{figure}
\centerline{\psfig{file=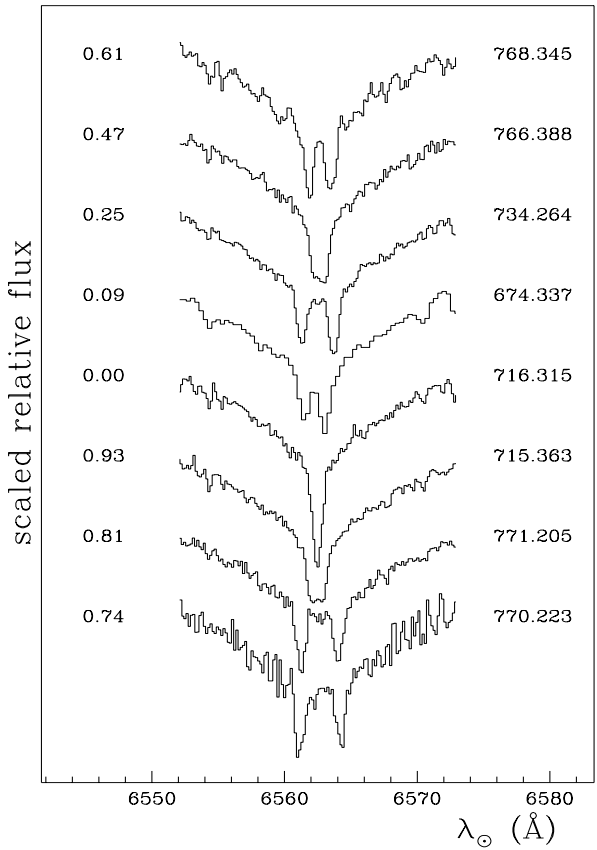,height=12cm,width=10cm}}
\caption[]{A sample of H$\alpha$ profiles of SS~Lac reckoned according to
the orbital phase. The merging and splitting of the spectral lines is
evident. Orbital phases on the left are according to Eq.(2).} 
\end{figure}

\section{Spectroscopic orbit}

Our observations were secured over a time span covering less than 8
revolutions of SS~Lac and therefore are not suited for a very accurate
period determination. A Deeming-Fourier analysis of data in Table~1 suggests
a period of 14.4$\pm$0.3 days, very close to the 14.416 days determined from
eclipse timing by several Authors (e.g. Brancewicz \& Dworak 1980). In the
computation of the orbit, the period 14.41638 from Eq.(2) has been adopted
(see below).

The orbital solution is given in Table~2. With the exception of $\Omega$ and
eccentricity, the formal errors for the other elements are $\leq$1.5\%, with
a weighted deviation of the observations from the computed orbit of
$\sim$0.8 km sec$^{-1}$. The solution is graphically presented in Figure~2.

It is worth to compare the photometric and spectroscopic orbital solutions
(cf. Savedoff 1951) through the combination of the eccentricity ($e$) and
longitude of the periastron ($\Omega$) elements (cf. Struve 1948):
\begin{equation}
e \,{\rm cos} \, \Omega = \frac{2 \pi}{P}\
\frac{(t_2 - t_1 - \frac{1}{2} \times P)}{(1 + {\rm cosec}^2 i)}
\end{equation}
The photometric value is +0.110 (Dugan \& Wright 1935), very close to the
+0.108 value from the spectroscopic orbital solution in Table~2 (and orbital
inclination from sect. 4.3). Thus the photometric and spectroscopic orbital
solutions appear in excellent agreement.

\begin{table}
\caption[]{Heliocentric radial velocities of the SS~Lac components.  Solid
line and filled circles in Figure~2 refer to star $a$. The quoted errors are
probable errors. When the two components were too close in radial velocity
for line splitting into individual components, a single value is given
(central column). {\sl MJD}$_\odot$ = JD$_\odot$ -- 2450000. }
\begin{flushleft}
\begin{tabular}{lrlrlrl}
\hline
       &         &      &&&           &         \\
MJD$_\odot$ & RV$_{\odot}^{a}$ & err. &  RV$_\odot$ & err.&  RV$_{\odot}^{b}$ & err. \\
       &         &      &&&           &         \\
659.537&    14.06&  0.21&&&    --49.99&    0.14 \\
660.562&    34.58&  0.39&&&    --68.13&    1.07 \\
673.414&         &      &--18.69&1.95&&         \\
674.337&    22.40&  1.61&&&    --54.14&    1.24 \\
675.361&    38.94&  1.41&&&    --72.88&    2.09 \\
701.448&         &      &--20.42&0.90&&         \\
714.336&  --73.89&  1.89&&&      35.91&    4.56 \\
714.378&  --71.98&  1.55&&&      33.63&    1.27 \\
715.329&         &      &--22.27&1.03&&         \\
715.363&         &      &--19.00&0.94&&         \\
716.315&         &      &--17.17&1.08&&         \\
716.561&         &      &--18.40&0.35&&         \\
717.530&    20.53&  0.28&&&    --58.77&    0.15 \\
732.271&    25.41&  0.38&&&    --65.65&    0.74 \\
732.317&    25.87&  0.09&&&    --67.03&    0.26 \\
734.264&    44.13&  1.01&&&    --83.81&    1.57 \\
748.391&    41.49&  1.04&&&    --85.81&    0.80 \\
765.510&    26.04&  2.00&&&    --70.85&    1.19 \\
766.209&     0.75&  2.10&&&    --60.55&    2.00 \\
766.388&     0.09&  3.45&&&    --47.79&    2.38 \\
767.326&         &      &--18.14&1.41&&         \\
768.345&  --64.94&  1.09&&&      24.18&    1.59 \\
770.223& --102.93&  0.50&&&      64.26&    0.50 \\
771.205&  --97.58&  0.50&&&      53.61&    0.50 \\
       &         &      &&&           &         \\
\hline
\end{tabular}
\end{flushleft}
\end{table}

\section{Discussion}

\subsection{Eclipse timing}

The {\sl MDJ}$_\odot$=716.315 spectrum in Figure~1 presents a perfect radial
velocity superposition of the two spectra, thus it corresponds to eclipse
conditions. The eclipse ephemeris given by Mossakovskaya (1993)
\begin{displaymath}
Min \ I \ = \ 2415900.76 + 14.41619 (\pm0.00013) \times E
\end{displaymath}
predicts a +0.456 day shift for the {\sl MJD}$_\odot$=716.315 spectrum,
which is comparable with the propagation of the uncertainty on the period. 
Imposing phase coincidence between Mossakovskaya's principal minimum and
{\sl MJD}$_\odot$=716.315 spectrum, this leads to the improved ephemeris
\begin{equation}
Min \ I \ = \ 2450716.32(\pm0.15) + 14.41638 (\pm0.00010) \times E
\end{equation}
The slightly longer orbital period nearly coincides with those given by
Brancewicz \& Dworak (1980, P=14.4163 days) and by Dugan \& Wright (1935,
P=14.41629 days).

\begin{table}
\tabcolsep 0.08truecm 
\caption{
Orbital elements for SS~Lac. Where appropriate, the second value correspond
to the component $b$ represented by a dotted line and open circles in
Figure~2. The quoted errors are the formal errors of the orbital solution.
The entry ``{\sl deviation}" is the mean weighted deviation of the observed
radial velocities from the computed orbital solution (weight = err$^{-2}$ in
Table~1).  \ {\sl MJD}$_\odot$ = JD$_\odot$ -- 2450000. The error on the
masses is $\sim$0.1 M$_\odot$.}
\begin{tabular}{lllll} \hline
&&&\\
                     && \multicolumn{1}{c}{star $a$} &&\multicolumn{1}{c}{star $b$}\\ \cline{3-3} \cline{5-5}
&&&\\
period               & (days)          & 14.41638           &&                   \\
baricentric velocity & (km sec$^{-1}$) & --21.2$\pm$0.3     &&                   \\
semi-amplitude       & (km sec$^{-1}$) & 74.7$\pm$1.1       && 77.6$\pm$1.4      \\
eccentricity         &                 & 0.122$\pm$0.019    &&                   \\
$a_i$sin$i$          & (AU)            & 0.098$\pm$0.001~~~ && 0.102$\pm$0.002~~~\\
$T_\circ$            & (MJD$_\odot$)   & 741.4$\pm$0.2      &&                   \\    
$\Omega$             & (deg)           & 332$\pm$9          &&                   \\       
deviation            & (km sec$^{-1}$) & 0.87               && 0.70              \\ 
&&&\\
inclination          & (deg)           & 78$\pm$5           &&                   \\
mass                 & (M$_\odot$)     & 2.80               && 2.69              \\
\hline
\end{tabular}
\end{table}

\subsection{Spectral classification}

The absence of He~I absorptions, the paucity of metallic lines and the very
strong Stark broadening of the wings of the Balmer lines suggest a spectral
classification around A2~V, virtually identical for the two members (cf.
Figure~1). A more refined classification however needs devoted observations
at the head of the Balmer series. Component $b$ appears slightly less
luminous than component $a$, as suggested by the relative depths of the
central Doppler cores and contribution to the overall H$\alpha$ line wings
in Figure~1. A lower brightness agrees with the slightly smaller mass of
component $b$ (cf. the semi-amplitudes and semi-major axes in Table~2) and
the just slightly reduced amplitude of $b$ eclipses compared to $a$ eclipses
(a few hundredth of a magnitude, cf. Dugan \& Wright 1935, Mossakovskaya
1993).
 
\begin{figure}
\centerline{\psfig{file=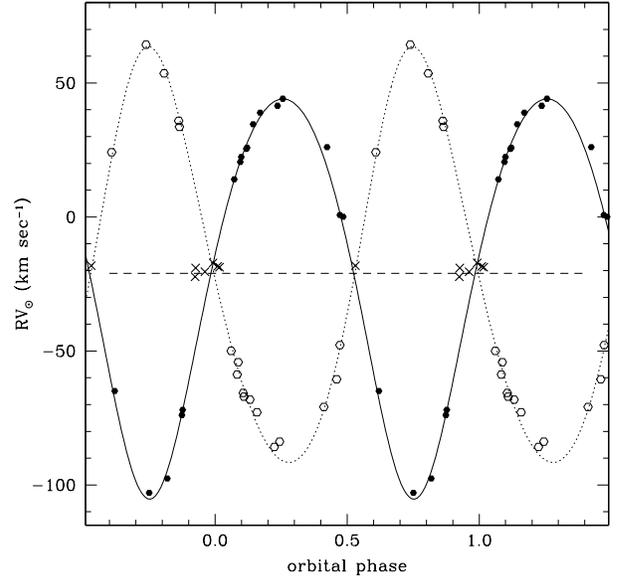,height=8cm,width=8cm}}
\caption[]{Orbital solution for SS~Lac. The curves give the orbital solution
of Table~2. Solid curve and filled circles refer to component $a$ in
Tables~1 and 2. The crosses mark spectra for which the too small velocity
separation caused merging of the two components into a single line. The
dashed line is the baricentric velocity (--21.2 km sec$^{-1}$).
Orbital phase according to Eq.(2).}
\end{figure}

\subsection{Orbital inclination}

The spectral classification of the two components can be turned into a
physical radius of R=2.25 R$_\odot$ from interpolation of tabular values
given by Allen (1973). The orbital separation from Table~2 is {\sl
a\,sin\,i} = 0.200$\pm$0.002 AU = 42.5$\pm$0.5 R$_\odot$. Lehmann (1991) has
stated that the system presented total eclipses (central eclipses given the
nearly identical size of the components) in 1900-1915, and afterward the
eclipses monotonically reduced in depth until they disappeared around 1960.
From separation and size of the two components it may be computed that at
the end of the eclipsing season the inclination was $i \sim 83^\circ$.  The
corresponding linear change in orbital inclination is
\begin{equation}
\frac{d{\rm i}}{dt} = 0.13 \pm 0.01 \ \ \ \ \ deg \  yr^{-1}
\end{equation}
which is significantly lower than the {\sl di/dt} = 0.18 deg yr$^{-1}$
obtained by  Lehmann (1991) on the assumption of a B7~V classification for 
both components (clearly ruled out by our spectra that show no evidence for
HeI absorption lines). From Eq.(3) and the occurrence of central eclipses at the
beginning of this century, the current orbital inclination is
\begin{equation}
i_{(1998)} \sim 78^\circ
\end{equation}
The above results indicate that the non-eclipsing season of SS~Lac lasts
for about 1275$\pm$90 years.

\subsection{Rotational velocities}

The rotational velocity of the components of SS~Lac has been derived by
comparison with standard stars from Sletteback (1975), which were observed
under the same instrumental conditions. The velocity turned out to be the
same for both components:
\begin{equation}
V_{rot} = 10 \ \pm \ 5 \ \ \ \ \ \ km \ sec^{-1}
\end{equation}
The dependence on inclination (see Eq[4]) has been removed under the
assumption that the rotational and orbital axes are parallel.  A $V_{rot} =
10\pm5$ km sec$^{-1}$ means that rotational and orbital motion are
synchronized or close to, the orbital velocity amounting to V$_{orb} = 8$ km
sec$^{-1}$.

It is worth noting that SS~Lac shows evidence for synchronization when the
orbital circularization is still far from being achieved. The two phenomena
seems therefore to evolve over quite different time scales, in agreement
with the theoretical expectations (cf. Zahn 1977).

The expected synchronization time for a binary with the SS~Lac parameters
may be estimated from the Tassoul's (1987) formalism to be:
\begin{eqnarray}
t_{syn}(yrs)& \geq & 2500 \times [P(days)]^{\frac{11}{4}}  \\
            & \geq & 4 \ million \ years  \nonumber
\end{eqnarray}
which is relatively short compared to the NGC 7209 cluster age (300 million
yrs according to Lynga 1985). The synchronization status cannot therefore be
used to decide if SS~Lac is or not a primordial binary of NGC 7209.

\subsection{Individual masses}

A straightforward application of the Kepler's III law to Table~2 data gives
2.69 and 2.80 M$_\odot$ for the masses of the two SS~Lac components.  The
tabular mass from Allen (1973) for the A2~V spectral type is 2.78 M$_\odot$,
in excellent agreement.

\subsection{Mass-Luminosity relation}

Vansevicius et al. (1997) have estimated a 1.0 kpc distance and a $A_V$=0.54
mag extinction to NGC 7209. Using a bolometric correction {\sl B.C.}= --0.29
mag for the spectral type A2~V (Allen 1973), the bolometric magnitude of the
SS~Lac components is
\begin{equation}
M_{bol} = +0.25 \ \ \ mag
\end{equation}
corresponding to a luminosity $L = 65 \ L_\odot$. The mean mass for the two
components is 2.745 M$_\odot$. At this mass the {\sl Mass-Luminosity}
relation has the numerical form (cf. Bowers \& Deeming 1984):
\begin{equation}
log \frac{L}{L_\odot} = 0.479 + 2.91 \ log \frac{M}{M_\odot}
\end{equation}
which predicts $L \sim 57 \ L_\odot$ for the SS~Lac components. The
agreement between the observed and predicted luminosity is satisfactory in
view of the uncertainties in the distance, reddening and the approximate
bolometric correction.

\begin{table}
\tabcolsep 0.08truecm 
\caption{Heliocentric radial velocity of the weak component of the Balmer lines
described in Sect.4.7. The errors may be quantified as $\pm$1 km sec$^{-1}$
when $N$=3, $\pm$2 for $N$=2 and $\pm$4 for $N$=1. {\sl MJD}$_\odot$ =
JD$_\odot$ -- 2450000. $N$ = number of Balmer lines used.}
\begin{tabular}{cclcclccl} \hline
&&&&&\\
MJD & RV$_\odot$&N~~~~& MJD &RV$_\odot$ & N~~~~ & MJD &RV$_\odot$ & N\\     
    &(km/sec)&&&(km/sec)&&&(km/sec)&\\
&&&&&&&\\
659.537 & --16 &  1  & 732.317 & --27 &  1 & 768.345 & --7  &  1 \\   
660.562 & --22 &  3  & 734.264 & --19 &  3 & 770.223 & --41 &  1 \\ 
675.362 & --29 &  1  & 748.391 & --23 &  2 & 771.205 & --19 &  1 \\
714.378 & --20 &  1  &         &     &    &         &     &    \\
&&&&&&&\\
\hline
\end{tabular}

\end{table}

\subsection{The unseen third body}

Lehmann (1991) ascribes the rotation of the plane of the orbit to the
presence of an unseen third body in the system, orbiting at a great distance
the closer central pair. A discussion of this hypothesis is beyond the
scopes of the present paper. We may however point out that in our spectra
when the Balmer lines show a clear split and the S/N ratio is high, a weak
absorption is visible. The H$\alpha$ component does not correspond to known
telluric absorptions in this region, and the coincidence between the various
Balmer lines favour an interpretation as a {\it real} stellar line. To the
benefit of future Investigators of SS~Lac we report in Table~3 the
heliocentric velocity of this weak component as measured on our spectra
(same weighting as described in Sect.2), that could in some way be related
to the unseen third body.

\acknowledgements{We would like to thank the referee, E.F.Milone, for accurate
reading of the manuscript and scrutinity of the data. We thank also
M.Rejkuba who secured a few of the spectra used in this investigation and
C.Barbieri, G.Cremonese and S.Verani who allowed us to secure a few other
spectra during their observing time. R.Margoni contributed useful insights
in the statistical treatment of the data.}

\end{document}